# Time and position sensitive single photon detector for scintillator read-out

**Sven Schössler**[a], **Benjamin Bromberger**[b], **Michal Brandis**[c], **Lothar Ph.H. Schmidt**[d], **Kai Tittelmeier**[b], **Achim Czasch**[a], **Volker Dangendorf**[b] **and Ottmar Jagutzki**[a*]

[a] *RoentDek GmbH,*
  *Im Vogelshaag 8, 65779 Kelkheim, Germany*

[b] *Physikalisch-Technische Bundesanstalt,*
  *Bundesallee 100, 38116 Braunschweig, Germany*

[c] *Nuclear Physics Division, Soreq NRC, Yavne 81800, Israel*

[d] *Institut für Kernphysik, Johann Wolfgang Goethe Universität*
  *Max-von-Laue-Str. 1, 60438 Frankfurt am Main, Germany*
  *E-mail:* jagutzki@atom.uni-frankfurt.de

ABSTRACT: We have developed a photon counting detector system for combined neutron and γ radiography which can determine position, time and intensity of a secondary photon flash created by a high-energy particle or photon within a scintillator screen. The system is based on a micro-channel plate photomultiplier concept utilizing image charge coupling to a position- and time-sensitive read-out anode placed outside the vacuum tube in air, aided by a standard photomultiplier and very fast pulse-height analyzing electronics. Due to the low dead time of all system components it can cope with the high throughput demands of a proposed combined fast neutron and dual discrete energy γ radiography method (FNDDER). We show tests with different types of delay-line read-out anodes and present a novel pulse-height-to-time converter circuit with its potential to discriminate γ energies for the projected FNDDER devices for an automated cargo container inspection system (ACCIS).

---

[*] Corresponding author.

# Contents

## 1. Introduction

Image intensifiers with micro-channel plates (MCP) are widely used for imaging at lowest light intensities and/or for counting single photons [1]. A special variant of this is the MCP-photomultiplier (MCP-PMT), which can detect single photons from red to UV wavelengths with a quantum efficiency of up to a few ten percent, providing spatial and temporal information with a precision well below 0.1 mm and 100 ps, respectively, at count rates up to a few million per second and for active detector diameters of at least 75 mm [2]. We have developed a novel design for a special MCP-PMT with so-called resistive screen anode (RS-PMT) adopting the technique of capacitive image-charge coupling through solid materials [3], which eases the mechanical design for position- and time-sensitive photon detectors and simultaneously allows reconfiguring the read-out method for various applications [4]–[6]. It especially enables implementing advanced read-out electronics for correlating each detected photon with coincidently recorded particles/photons on auxiliary detectors or with external parameters. Such a correlated "event" contains information not only on the position and arrival time of each photon detected with the RS-PMT but also on relevant physical parameters of the correlated particles detected by a greater parent detection device in a complex experiment. All these correlated parameters are then individually "listed" for each registered photon. In this work we demonstrate the features of such an event-counting image intensifier (ECII) for an application of simultaneous neutron and γ radiography.

It has been demonstrated that an RS-PMT can be a powerful tool for imaging of photons from scintillators, e.g. generated by nuclear particles or X/γ-rays [7]. An example is the method of Fast Neutron Resonance Radiography (FNRR), utilizing a pulsed beam of fast neutrons with continuous energy spectrum (typically 1–10 MeV) to investigate the composition of bulky objects. A time- and position-sensitive detector placed downstream is registering the transmitted neutrons. The energy of each neutron is determined by measuring its time-of-flight (TOF) from the production target to the detector and stored with the corresponding position coordinates digitally in a "list-mode" file. Off-line data analysis of the latter allows generating a set of images from this list for certain neutron energy windows. Thus, utilizing the element and neutron-energy specific total neutron cross-section structures allows identifying the elemental composition of a sample (see Figure 1**:** ).

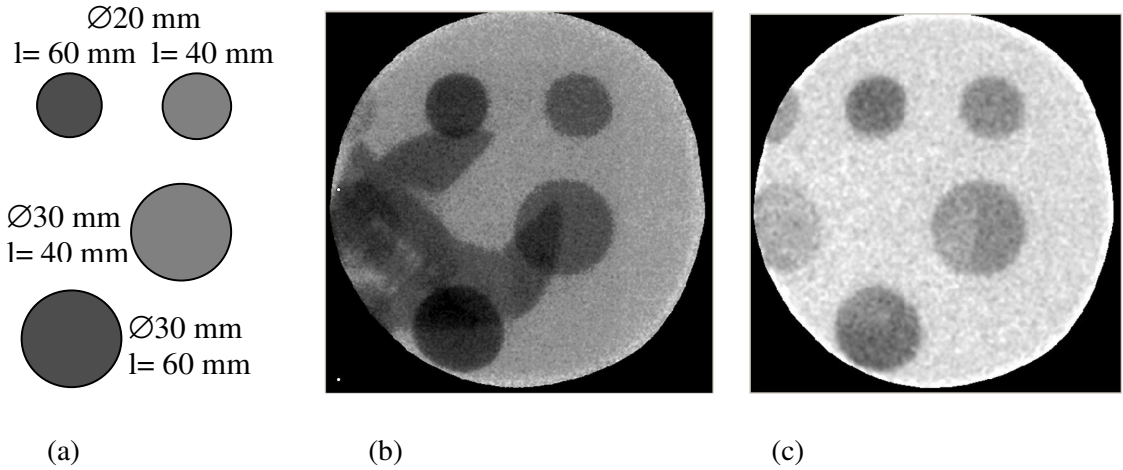

(a)            (b)            (c)

**Figure 1:** Neutron images of a sample formed by various carbon rods and a steel wrench. (a) shows the configuration (length *l* and diameter ∅ of the carbon rods). (b) shows the radiographic image in a broad-energy neutron beam (1 – 10 MeV) and (c) the carbon distribution after utilizing the FNRR technique. The example is taken from reference [2].

In order to investigate the feasibility of a projected automated cargo container inspection system (ACCIS) for contraband detection including nuclear material [8]-[11] several detector techniques are currently tested. One detector concept is based on a fast gated multi-frame CCD camera [11]. This concept offers a limited number of simultaneously measured energy frames (currently up to 8) but at potentially very high neutron fluxes. The alternative concept, TRECOR (from "Time-Resolved Event Counting Optical Radiation Detector") [12], is more limited in rate but offers a practically unlimited number of energy groups simultaneously. TRECOR employs a single particle counting method as described here. With the TRECOR approach the FNRR method can be amended by a simultaneous Dual Discrete Energy γ-Radiography (DDER) technique. A prerequisite for such a combined Fast Neutron Dual Discrete Energy γ-Radiography (FNDDER) technique is a detection system which is able to discriminate neutron counts from those of γ-rays, determine neutron energy by TOF measurement as well as the γ energies, by means of measuring the intensity of the scintillation photons' flash. Since expected neutron and γ fluxes are of the order of a few times $10^6$ $cm^2 s^{-1}$, both the analogue electronics and the data acquisition devices must be able to operate at very high rates. In this work we show tests of an RS-PMT detector and read-out components optimized for high-rate operation and introduce a very fast pulse-height analyzing electronic device which converts pulse height (i.e. energy) into time information that can be easily implemented into the fast data acquisition flow.

## 2. Detector system for Fast Neutron Dual Discrete Energy γ-Radiography (FNDDER)

The design of the detection system for the ACCIS project follows an earlier experimental setup for FNRR at the Physikalisch-Technische Bundesanstalt in Braunschweig (PTB) [2]. Figure 2: shows a schematic view of the detector.

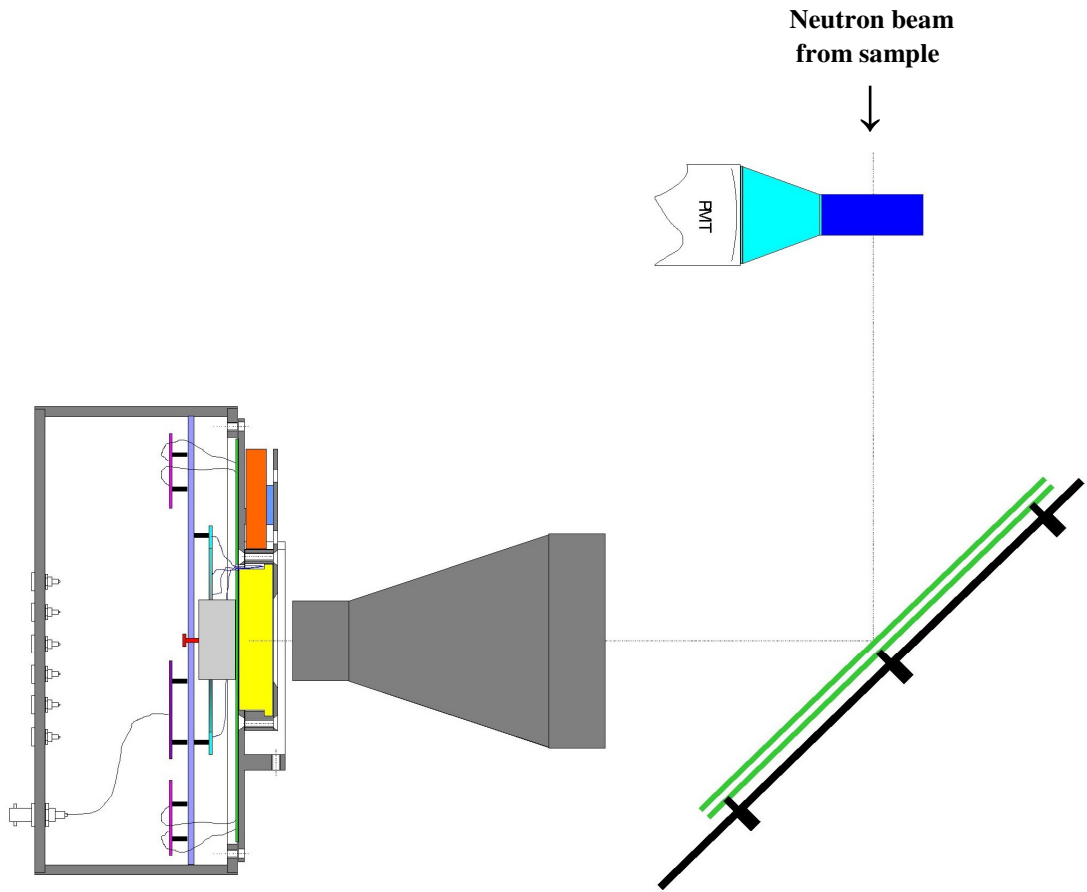

**Figure 2:** Sketch (above) and photograph (below) of the electro-optical elements in the FNDDER setup. An updated and more detailed presentation of the detection system can be found in this proceedings [12].

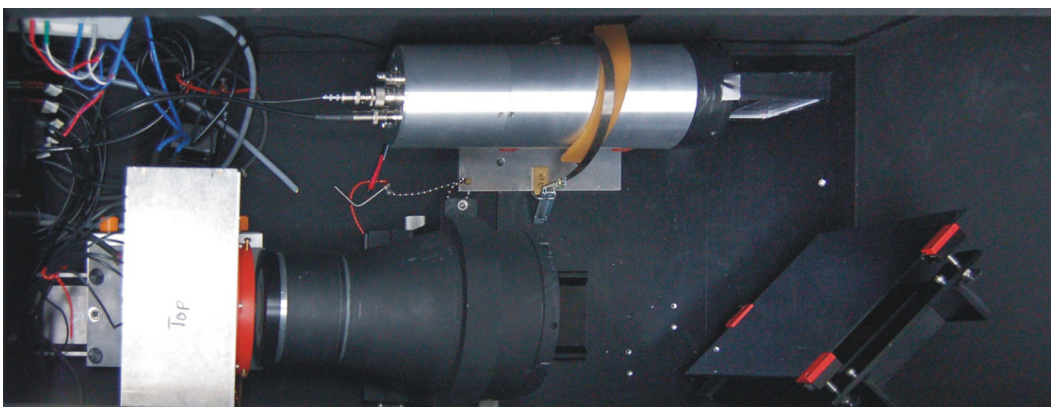

The incoming radiation is converted in a plastic scintillator (for neutrons) or a high-Z anorganic scintillator (for γ-radiation). A standard PMT is coupled to the side of the scintillator by a perspex taper and collects some of the scintillation photons created by a neutron or γ-ray. A downstream mirror reflects light from the scintillator towards a large-aperture lens backed by the RS-PMT as image sensor. The RS-PMT is equipped with an external delay-line anode for retrieving the spatial (and also temporal) information from picking up the local image charge. The standard PMT is capable of fast event timing and collects scintillation light with much higher efficiency than the lens/RS-PMT branch. This is used to measure the magnitude of γ-induced photon showers in the scintillator to determine the energy deposited by radiation and to evaluate the original energy of the detected γ quant. The advanced analogue and digital read-out electronics is finally correlating all detection elements and forms the FNDDER system. In the following we will describe the RS-PMT with read-out anode for high throughput and the novel pulse-height converting read-out electronics in greater detail.

## 2.1 The RS-PMT with Hexanode read-out for high throughput

The general operation principle of the RS-PMT with delay-line read-out anode has been described in detail before [2], [13]. In short, an RS-PMT is very similar to an MCP-PMT used in advanced night vision devices. Only the phosphor-coated rear window is replaced by a ceramic wall with a resistive coating ("resistive screen"). The screen resistivity is chosen so that nearly 100% of the collected charge from the MCP avalanche (triggered by single photon/particle impact) is capacitance coupled as image charge onto a metal electrode ("anode") placed on the rear wall outside the sealed tube, see Figure 3:. To stress the analogy further: the capacitive coupling can be seen as the pick-up of microwaves passing through the ceramics to the air-side anode, much like light from a standard MCP-PMT's phosphor screen is transmitted through a glass window and recorded on a CCD/CMOS chip. Due to the similarity of the mechanical design the RS-PMT is as suitable for mass production as standard MCP-PMTs are.

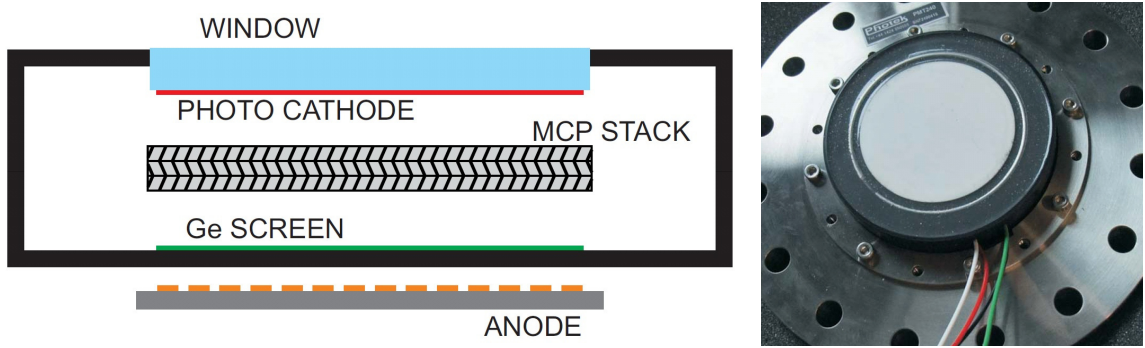

**Figure 3:** left: sketch of an RS-PMT for single photon counting and imaging. The photo-cathode on the glass entrance window, a high-gain triple MCP stack and a highly resistive screen coating, e.g. made by a thin film of Germanium on a ceramic rear wall are inside a sealed PMT vacuum housing. The read-out electrode (anode) is placed outside near the rear wall (usually in contact with it). There, the image charge is naturally spread to a few mm footprint size allowing for high resolution center-of-mass averaging with read-out anodes like the delay-line. Right: rear view of a 40 mm active RS-PMT (here: mounted on a flange).

The pickup electrode (anode) behind the RS-PMT is a standard printed-circuit board (PCB) which can be seen as the first stage of a greater "electronic camera" system for the RS-PMT with electronic modules processing the signals towards a digital image. One advantage of the RS-PMT approach compared to MCP-PMTs with enclosed anodes [14] is that the read-out can easily be reconfigured by replacing the pick-up anode and the electronics behind it. From the different electronic camera concepts we favor the delay-line read-out due to its ease, intrinsic charge centroiding and short dead-time. The delay-line anode encodes the mean position of the detected image charge into a time sequence of signals picked up from terminals of adequately structured electrodes.

By measuring the time sequence of these signals utilizing fast amplifiers, precise timing ("constant fraction") discriminators (CFD) and a multi-fold time-to-digital converter (TDC) the position and absolute arrival time (i.e. the TOF) of each particle/photon can be registered and stored at MHz count rates. The delay-line anodes used behind the RS-PMT in this work are standard multi-layer PCBs consisting of rows of diamond-shaped pads connected along one direction. These rows are interconnected on one end either via serpentine tracks (meander delay-line) or a network of coils and capacitors (LC delay-line), see Figure 4:. Both methods introduce a certain delay for induced signals after emerging from the rows and travelling along the so-formed delay-lines in both directions towards the terminals. Since the image charge is spread over more than one row, center-of-gravity averaging allows for a spatial precision not limited by the row-to-row distance (anode pitch). At least two of such layers are needed for a two-dimensional detection of the incoming particle or photon. Unlike anodes for direct charge collection, the layers for each dimension can be separated by a continuous insulating sheet.

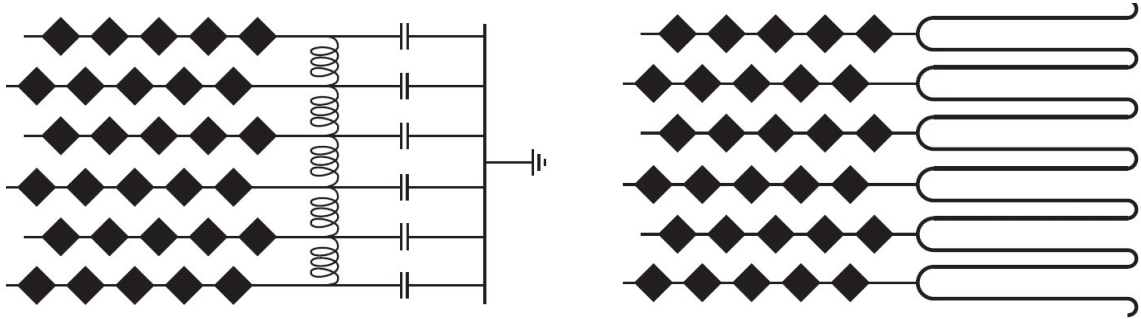

**Figure 4:** Delay-line layers for RS-PMT read-out. Rows of interconnected pads pick up the image charge signal which is then guided along delay-lines to terminals for further signal processing (amplification, timing discrimination and digitization). Right: delay-line formed by a meandering track, left: delay-line formed by a network of coils and capacitors.

In the earlier FNRR experiments we have used a two-layer anode with LC delay-lines. For the FNDDER we also investigate meander delay-lines because of their advantages at high throughput. Signals traveling along a properly dimensioned meander-shaped transmission line are less broadened and therefore have a higher signal-to-noise ratio. This allows reducing the MCP gain which is inevitable at very high rates and also increases the lifetime of the RS-PMT. Low-damping meander lines have to be kept comparably short and have a lower delay per pitch. Although this may compromise the spatial resolution it is anyway necessary for reducing the dead-time of the anode, i.e. the time interval for a signal to pass across the delay-line. The latter is also the main reason why we use three-layer anodes, so-called Hexanodes (see Figure 5: ) for the FNDDER. Although these anodes increase the number of electronic channels needed and the amount of transferred data per event by 50 %, a Hexanode offers the big advantage that it can still operate without losing events at rates when - due to Poisson statistics - more than one particle is detected within the dead-time of the anode [15]. Another advantage of the Hexanode is an intrinsic control and calibration of the image linearity [16].

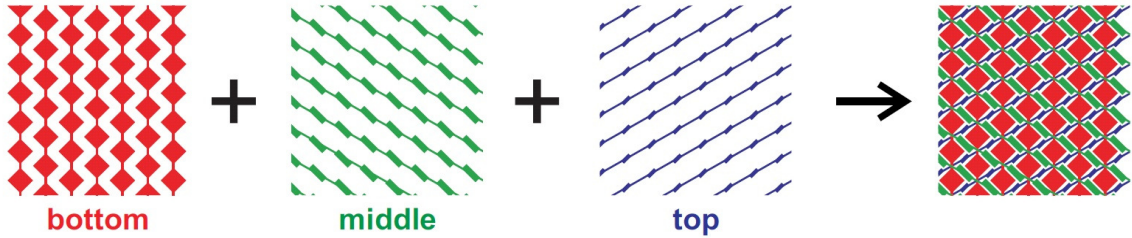

**Figure 5:** Above: structures and orientation of the pick-up pad rows for a three-layer Hexanode geometry. The pad sizes in the multi-layer PCB grow as layers are located further away from the resistive screen. This compensates for the diminishing image charge on more distant layers. Below: LC delay-line Hexanode mounted on an RS-PMT (left) and meander delay-line version (right). The transmission lines are covered between ground planes in the multi-layer PCB.

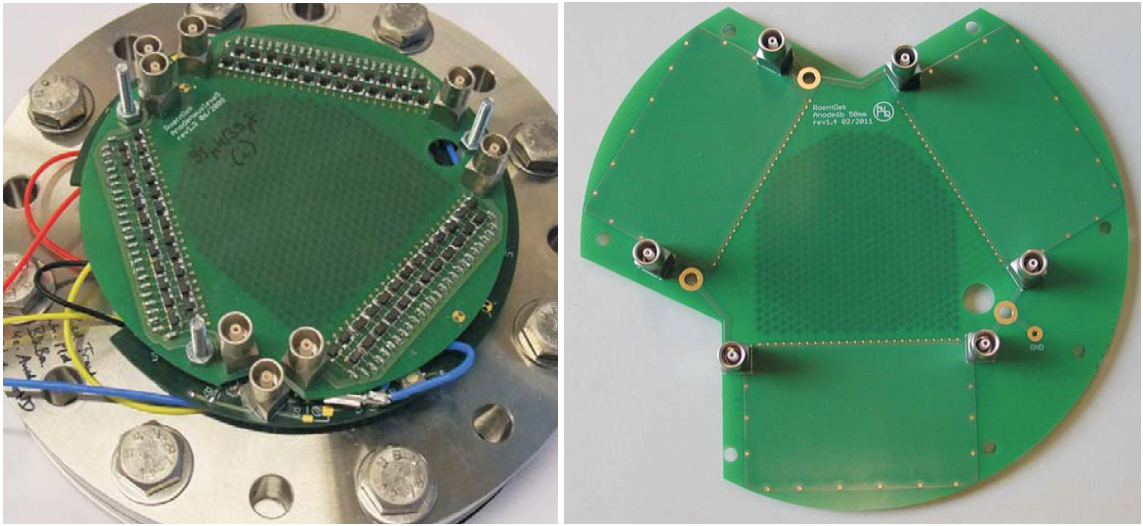

Several LC and meander delay-line geometries for the Hexanode have been tested for different path delays. The spatial resolution was found to be 0.1 mm FWHM or better (see Figure 6: ) which is more than sufficient for FNDDER. However: this high spatial precision can only be maintained as long as the signal-to-noise ratio is sufficiently high. Generally, an MCP stack in an RS-PMT optimized for high rate operation in single counting mode can provide a gain up to a factor of 10 million at rates up to 1 MHz. This provides a spatial resolution below 0.1 mm. If the rate shall be further increased the pulse height will drop, yielding a maximum gain of only 1 million at about 10 MHz [17] and inferior spatial resolution. Furthermore, it may generally be required to operate the detector at low gain for prolonging its lifetime.

Tests at low gain operation are still due and will be continued with the next RS-PMT prototype. Optimal parameters for the FNDDER application in terms of expected count rate, spatial resolution demand and projected lifetime are still to be investigated. Only after that the optimal read-out anode geometry can be chosen.

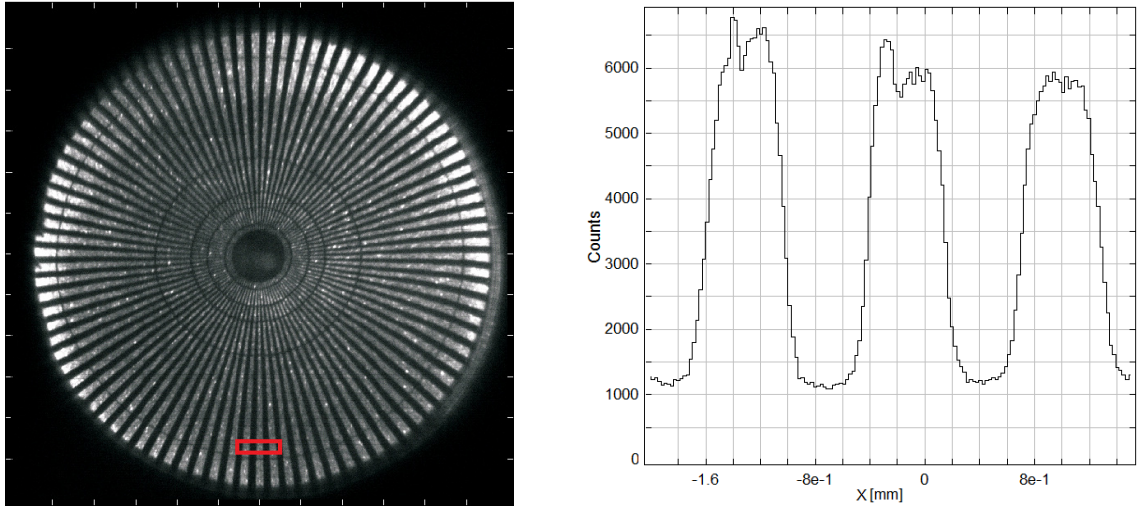

**Figure 6:** Left: Shadow image of a 40 mm test mask. Data within the red rectangle are projected to the x-axis (right picture). Right: Line scans allow an estimation of the spatial resolution being around 0.1 mm FWHM both for a LC delay-line (1.5 ns delay per mm) and a novel meander delay-line with only 0.75 ns/mm.

## 2.2 The CFDx timing discriminator device with pulse-height-to-time converter

The electronic read-out concept of the existing FNRR system based on the RS-PMT already had the capability of being operated at MHz count rate. It also could read out several detectors in coincidence (i.e. the PMT and RS-PMT) for retrieving spatial and temporal information with high precision. The remaining task was to include a pulse-height (energy) measurement for the γ-ray-generated signals into the data acquisition stream. The problem is that standard pulse-height analyzing circuits have a high dead-time of the order of several μs. Even if this could be reduced by compromising on the energy resolution (the requirements for this application are comparably low) it remains difficult to include this information into the digital data stream at MHz rates.

Therefore it was decided to design an analogue pulse-height-to-time converter circuit (PHTC) which codes different pulse heights into a proportional time delay between two standard signals (here: NIM-pulses): Thus, the delay between pulses is, in first order, a linear function of the pulse height which again is a measure of the energy deposited in the scintillator by the radiation quant. The transition times of these signals are then recorded using free channels of the same TDC unit which already records the time-coded position and TOF information of the same event. The additional parameter increases the amount of processed data only by 10-20% and a correlated recording of the energy coordinate is intrinsically guaranteed. The dead-time of the analogue PHTC circuit needs to be of the order of a few 10 ns (like for the signals from the delay-line anode) for not limiting the throughput. Since the application requires only differentiating between a low (4.4 MeV) and a high (15.1 MeV) γ-line the energy resolution and linearity requirements are not very demanding: about 10 % FWHM resolution and integral linearity of 30% over the relevant pulse height range are fully sufficient.

For this task a new dedicated module, the CFDx, was designed, see Figure 7: . A novel "ramp" PHTC circuit was incorporated into the CFD unit already used for determining the TOF. The analogue input signal is internally multiplexed to circuits for analyzing the signal timing (standard CFD function) and the ramp PHTC circuit which turns the trailing edge of the input signal into a stretched and almost linearly decaying slope. After a certain level is reached the ramp comparator triggers a digital output (NIM signal): the higher the input signal the later comes the trigger relative to the CFD's timing signal. It is even possible to code the timing and the pulse height of an analogue input signal into a single NIM signal: its leading edge refers to the TOF, while the length of the signal is a linear function of the analogue pulse height, i.e. the energy. Thus a single TDC channel with dual-slope detection capability may be used both for TOF and pulse-height measurement. The dead-time of the PHTC circuit is only 20-30 ns depending on the chosen slope of the ramp. In case of a fairly steep slope (lowest dead-time) the pulse-height resolution is well below 5 % in the range from 10 % to 90 % of full scale, even when effects of integral and differential non-linearity are not corrected for, see Figure 8: . This makes the new device (called CFDx) a more than adequate pulse-height analyzer for the FNDDER system, see Figure 9: .

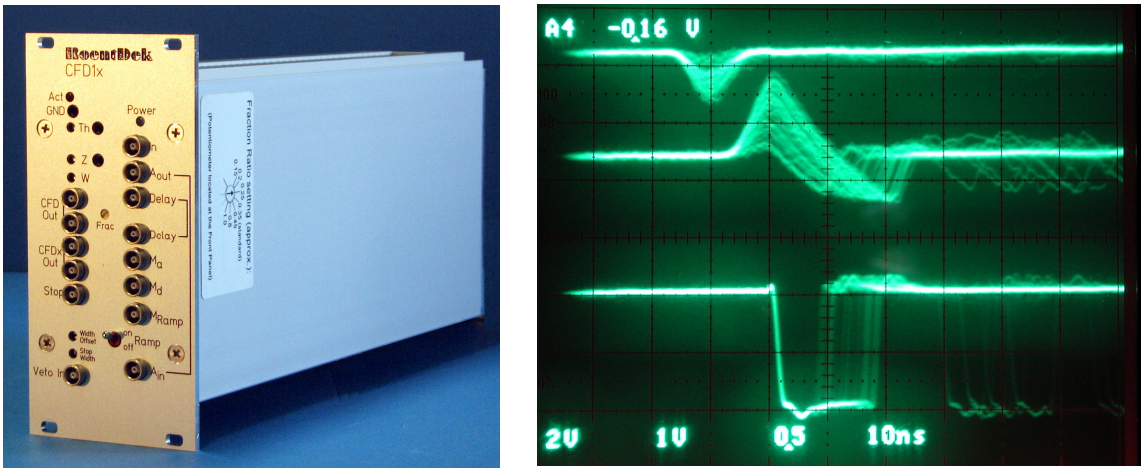

**Figure 7:** CFD1x module containing the CFDx circuit in a 19" 3HU case (left). Right: oscilloscope signal traces of the CFDx output demonstrating its function. Upper trace: input signals with different pulse heights, middle trace: signals from the "ramp monitor" output. Lower trace: NIM output signals with varying lengths as a function of the input signals' pulse heights.

### 3. Tests of the FNDDER detection system

In the projected FNDDER method for ACCIS the sample will be illuminated simultaneously by a fast broad-energy neutron beam probing the transmission at various neutron cross-section resonances (revealing the detailed composition of low-Z materials) and by γ-rays of two discrete, well separated energies (4.4 and 15.1 MeV) for high-contrast radiography in the high-Z range. The pulsed broad-energy neutron beam and the dual discrete-energy γ-ray beam are produced in a $^{11}$B target by a pulsed deuteron beam in the same nuclear reaction $^{11}$B(d,n+γ)$^{12}$C. The neutrons can easily be separated from the γ-component by the TOF method, the same technique used to measure the neutron energies.

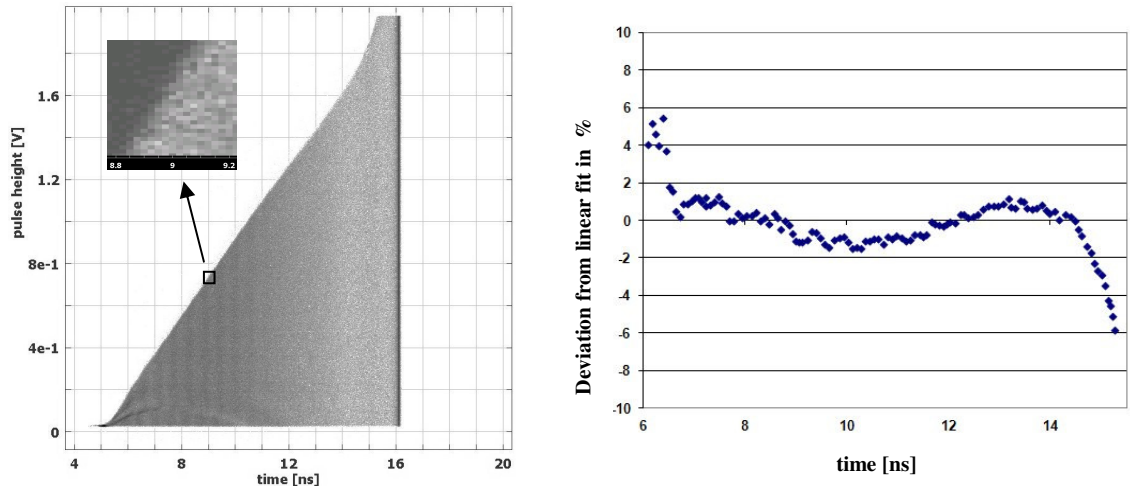

**Figure 8:** Response function of the CFDx. The 2d plot (left) shows measured pulse heights of signals as function of an upper pulse-height gate. Ideally there should be a smooth distribution of continuously decaying intensity (grey scale) from left to right, according to the input signals' pulse-height spectrum. The diagonal boundary would follow a straight line in case of perfect linearity response. Deviations from that describe the linearity function (right picture). The undulation in this function (also visible as undulating shading in the left picture) is due to differential non-linearity. The inset with inverted colour code (left) shows the definition of the boundary which is a direct measure of the resolution: Intensity drops from 90% to 10% within 20 mV (or 0.1 ns, respectively). This corresponds to a resolution of 1% of full scale.

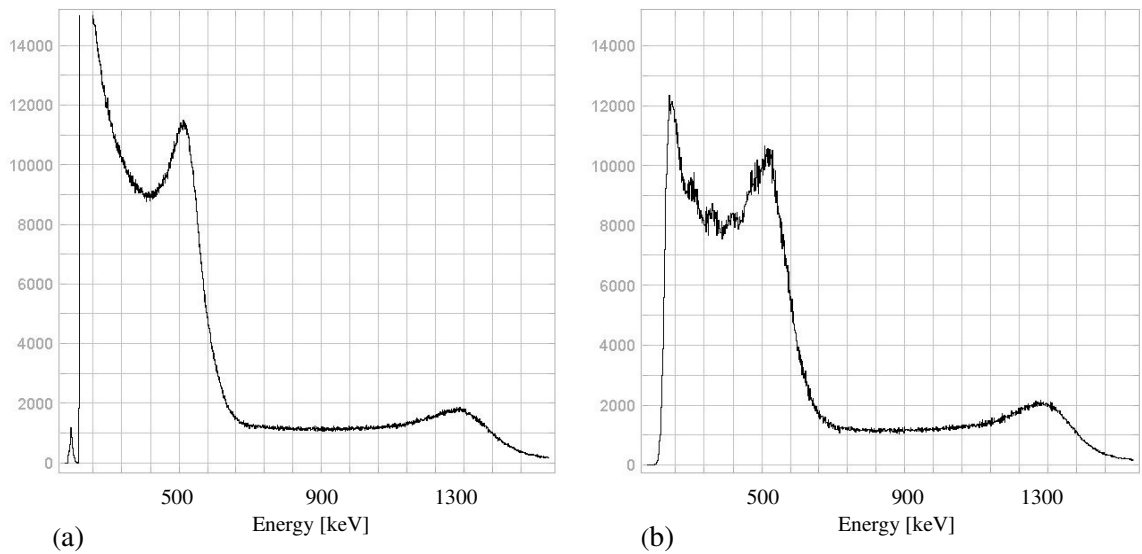

**Figure 9:** Comparison of the $^{22}$Na γ spectrum detected by a NE213 scintillation spectrometer, (a) with standard charge-integrating electronics and (b) with the CFDx module described here (right). The deviations are due to different threshold settings and statistics (y-scales normalized at 511 keV line) during the measurements and due to slight non-linearities in the CFDx response (no linearity correction function has been implemented here). The differential non-linearity contributes to the undulation near the 511 keV line in (b).

So far the events in the "Gamma peak" have been rejected in the data analysis as "background" and only the neutron events were analyzed for producing element-sensitive images as in Figure 1:. In FNDDER the γ peak would contain (besides some background from frame overlap, i.e. neutrons from the previous beam burst) all γ energies but predominantly the two lines (4.4 and 15.1 MeV) relevant for the dual-energy γ-radiography technique. These two γ energies can only be distinguished by measuring the deposited energy of the quant, i.e. the pulse height. First FNDDER test experiments were performed at the fast neutron facility of the PTB with the setup shown in Figure 2: and using the CFDx read-out electronics. Here, the detector was equipped with a LYSO scintillator which is used in our setup to measure the gamma image and spectrum. Figure 10 shows a two-dimensional intensity plot in logarithmic scale. All registered events from the PMT are displayed in a 2d histogram as function of TOF (x-axis) and pulse height (y-axis). Not shown are the position (and redundant TOF) data from the RS-PMT which have also been recorded in coincidence for each event.

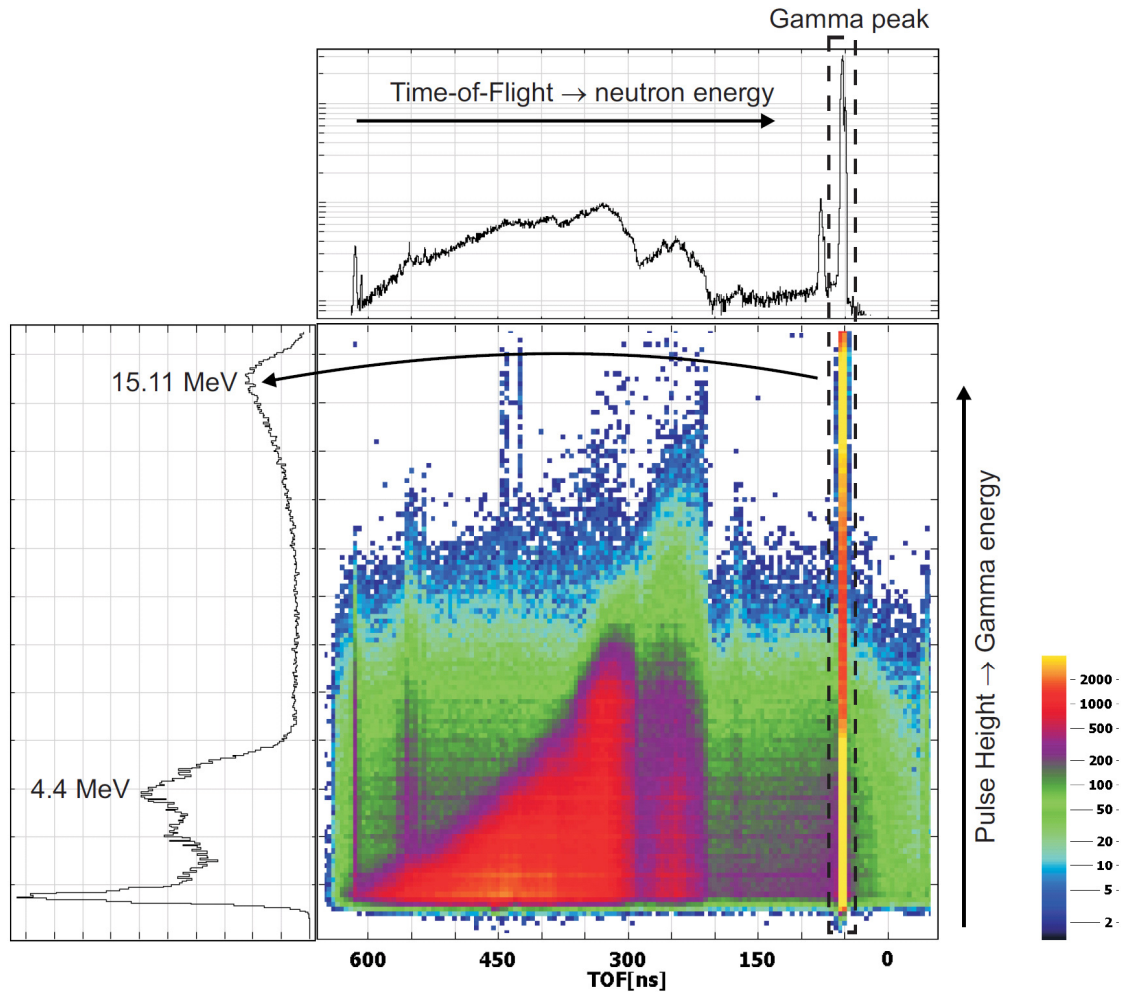

**Figure 10:** First tests of the FNDDER detection system with CFDx. The 2d plot shows the intensity distribution (color coded in log scale) for events of scintillation flashes measured by a PMT attached to a LYSO scintillator (see Figure 2: ) as functions of pulse height (corresponding to deposited energy) on the y-axis and the TOF of the neutrons and γ-particles on the x-axis (note that the TOF increases from right to left, therefore the neutron energy, which behaves with $1/TOF^2$, increases from left to right). The scintillator was irradiated by neutrons and γ-rays from a $^{11}B(d,n+\gamma)^{12}C$ reaction at the PTB accelerator facility.

The one-dimensional plot above the color plot shows a projection of all data to the x (TOF) axis in log scale. The γ peak can clearly be distinguished. All other events are produced from neutrons which react mainly with the Si in the LYSO via the (n,p) reaction, having a threshold of about 4 MeV. Setting a software gate on the γ peak leaves only the events produced by γ-rays. Pulse-height analysis for these events reveals the γ-energy spectrum as needed for DDER (Figure 10, left 1d plot). Therefore, the detected events can be attributed to a certain neutron or γ energy (with some ambiguity due to the low-energy fraction of the 15.1 MeV Compton background in the range of the 4.4 MeV line). A more detailed discussion of these results can be found in this proceedings by M. Brandis *et al.* [12].

## 4. Conclusion and Outlook

The test experiments have shown that the implementation of the CFDx circuit in the combined neutron/γ spectroscopic imaging setup, using a PMT and RS-PMT camera device for imaging light from a scintillator, can indeed measure the Compton spectrum and easily separate the two γ lines, which is the minimum requirement in the projected FNDDER system. All elements in the systems are optimized for operation at highest throughput, i.e. the maximum single particle/photon count rate achievable with nowadays' image intensifier technique. The next stage will be demonstrating the FNDDER performance in element-sensitive imaging for samples containing a variety of low-Z and high-Z materials. Another important feature to be determined is the maximum throughput up to which a certain image contrast can be maintained while signal-to-noise degrades due to physical limitations of the image intensifier or due to the need for low-gain operation for increasing lifetime towards practicable and commercially sensible periods.

### Acknowledgments

This work was supported by the German ministry of research and technology (BMBF) through a project grant in the framework "Forschung für die zivile Sicherheit", Förderkennzeichen 13N11154.